\documentclass[runningheads]{llncs}
\usepackage[T1]{fontenc}
\usepackage{graphicx}
\usepackage{pdfcomment}
\usepackage{xcolor}

\usepackage{footnote}
\usepackage{amssymb}
\begin{document}
\title{ICE-T: A Multi-Faceted Concept for Teaching Machine Learning}

%
%
\author{Hendrik Krone\inst{1,2} \and
Pierre Haritz\inst{1,2} \and
Thomas Liebig\inst{2,3}
}
%

\authorrunning{Krone et al.}
%
\institute{\textit{Both authors contributed to this work equally.}\\ 
\and TU Dortmund University, Dortmund, Germany \\
 \and
Lamarr Institute for Machine Learning and Artificial Intelligence, Dortmund, Germany\\
\email{firstname.lastname@tu-dortmund.de}
}
\maketitle              
\begin{abstract}
The topics of \textit{Artificial intelligence (AI)} and especially \textit{Machine Learning (ML)} are increasingly making their way into educational curricula. 
To facilitate the access for students, a variety of platforms, visual tools, and digital games are already being used to introduce \textit{ML} concepts and strengthen the understanding of how \textit{AI} works.
We take a look at didactic principles that are employed for teaching computer science, define criteria, and, based on those, evaluate a selection of prominent existing platforms, tools, and games.
Additionally, we criticize the approach of portraying \textit{ML} mostly as a black-box and the resulting missing focus on creating an understanding of data, algorithms, and models that come with it. 
To tackle this issue, we present a concept that covers intermodal transfer, computational and explanatory thinking, \textit{ICE-T}, as an extension of known didactic principles.
With our multi-faceted concept, we believe that planners of learning units, creators of learning platforms and educators can improve on teaching \textit{ML}.


\end{abstract}
\section{Introduction}
\textit{Machine Learning} (ML) is currently finding its way into more and more curricula due to its increasing significance.
Despite the availability of initial materials for classroom instruction, there is still a shortage of suitable resources for effective teaching.
In recent years, we have seen new platforms, tools, and games address the topic of \textit{Artificial Intelligence} (AI) in early education (primary school to high school). 
Studies have demonstrated that digital game-based learning can boost students' motivation and engagement, enrich their cognitive and emotional development, and thereby enhance their learning efficiency \cite{anastasiadis2018digital}.
Using digital games for teaching \textit{ML} offers the advantage that students can learn and experiment in a playful manner.
This not only enhances their comprehension of these complex subjects but can also spark their interest in the topic.

Visual tools are a popular way to teach \textit{ML} concepts. 
The variety of representations plays an important role in enhancing the learning experience. 
In addition, these formats can help to include diverse teaching methods, making complex concepts more accessible to a wider range of students.

Learning platforms play a crucial role in supporting educators.
There is currently a significant gap in teacher training for \textit{AI} skills \cite{SPERLING2024preserviceteacher}, indicating that many computer science teachers are unfamiliar with \textit{ML} techniques and find it challenging to teach them.
Using learning platforms enables them to access resources and materials tailored to the needs of students, contributing to more effective \textit{ML} education.

For education, various guidelines, frameworks and models exist. The Technological Pedagogical Content Knowledge (TPaCK) framework describes the knowledge required by educators to create technology-enhanced learning environments, emphasizing the interplay between content knowledge, pedagogy, and technological literacy \cite{Mishra2006}.  DPaCK strengthens digital literacy through subject-specific content, while AI-PaCK provides a structured description of AI teacher education \cite{Lorenz2023}. But didactic principles have not been sufficiently explored to develop a comprehensive and systematic approach for creating new tools, games, and platforms dedicated to teaching \textit{ML} content.
To fill this gap, we have answered the following research questions as our main contribution in this paper:
\begin{enumerate}
    \item \textit{RQ1}: To what extent do existing games, digital tools and platforms for teaching machine learning implement the facets of intermodal transfer, as well as computational and explanatory thinking? 
    \item \textit{RQ2}: How can a process model for teaching Machine Learning be derived from a Data Mining process model?
    \item \textit{RQ3}: How can the presented didactic principles be combined and applied? 
\end{enumerate}

To answer these questions, we first introduce three facets of thinking and the related work on the topic. 
Then, we compare the tools, games, and platforms and take a close look at the process model for Data Science. 
Finally, we present our proposal, \textit{ICE-T}, a multi-faceted concept for teaching ML.

\section{Background \& Related Work}
In this section, we will give an introduction to three facets of thinking and present other related work.
\subsection{Intermodal Transfer}
Intermodal transfer refers to the ability to apply knowledge or skills learned in one modality or context to a different modality or context. It involves the transfer of learning from one domain to another. In the context of education and cognitive development, intermodal transfer can be observed when individuals successfully apply what they have learned in one mode (e.g., visual, hands-on) to another mode (e.g., symbolic, abstract).

\begin{figure}[!hbt]
    \begin{minipage}{.45\textwidth}
            \centering
            \includegraphics[width=0.85\columnwidth]{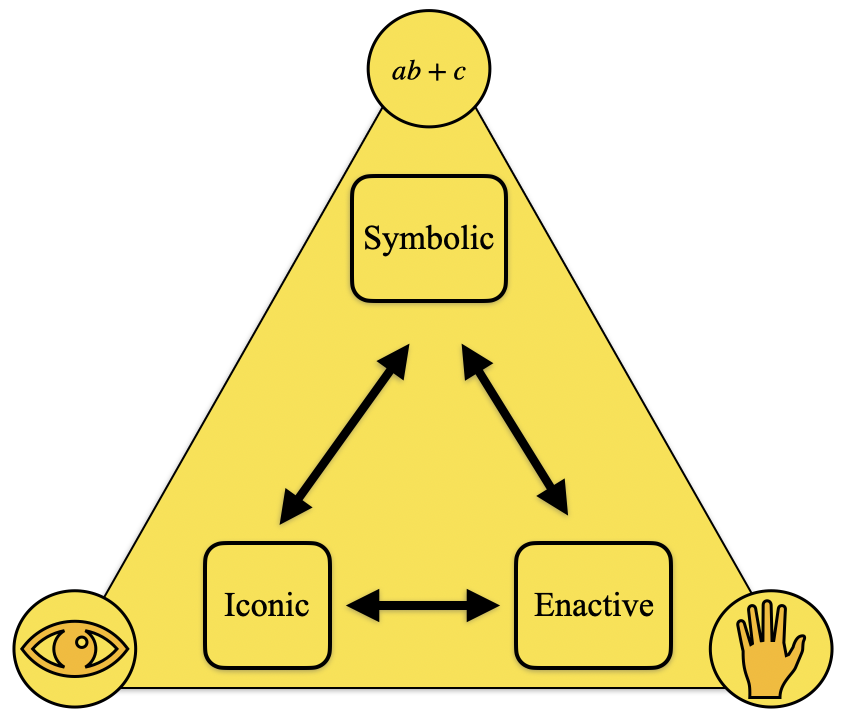}
            \caption{Intermodal Transfer via the EIS principle: Enactive - Iconic - Symbolic.}
            \label{fig:eis}
    \end{minipage}
    \hspace{.1\textwidth}
    \begin{minipage}{.45\textwidth}
        \centering
        \includegraphics[width=0.85\columnwidth]{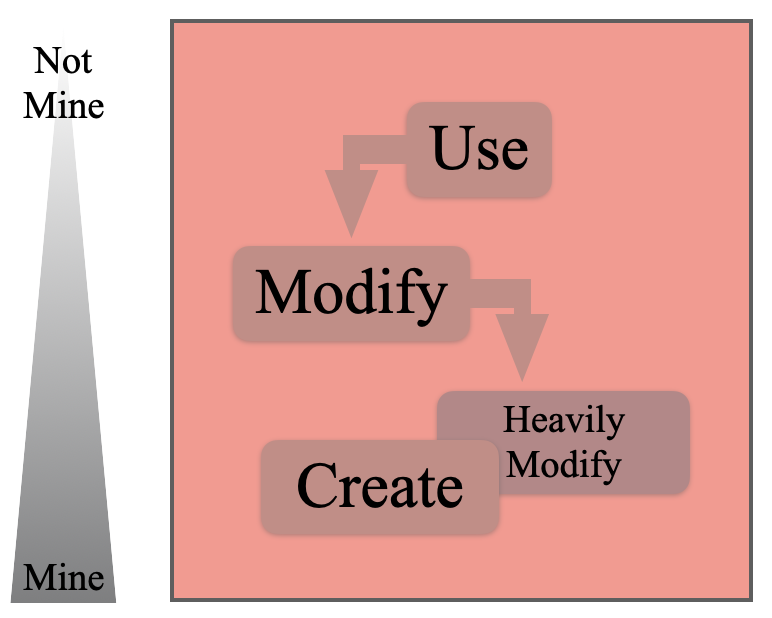}
        \caption{Computational Thinking via UMC: Use - Modify - Create.}
        \label{fig:umc}
    \end{minipage}
\end{figure}

The principle of \textit{EIS} (Enactive, Iconic, Symbolic) by Bruner \cite{Bruner1978} is a framework used in educational psychology, specifically in the context of learning and cognitive development.
The principle is visualized in figure \ref{fig:eis}. This framework describes three levels of representing information that learners advance through as they acquire knowledge and skills. The \textit{EIS} principle is closely related to the \textit{spiral approach}, in which students progress through a subject with ever-increasing complexity \cite{Bruner1978}. In the case of the \textit{EIS} principle, intermodal transfer can be evident in the progression from enactive to iconic to symbolic representation. Learners develop the ability to transfer knowledge and skills across different modalities, for example:
\begin{itemize}
    \item \textit{Enactive to Iconic:} Learners transfer knowledge gained through physical experiences (enactive) to mental images or visual representations (iconic). For example, a child who has learned to balance on a bicycle (enactive) can transfer this knowledge to create a mental image of bike riding.
    \item \textit{Iconic to Symbolic:} As learners move from iconic to symbolic representation, they transfer their understanding from visual or sensory images to abstract symbols or linguistic forms. This involves the ability to connect concrete visual representations to more abstract and conceptual symbols.
\end{itemize}

The intermodal transfer perspective emphasizes the flexibility and adaptability of cognitive processes, highlighting that learning in one modality can enhance performance or understanding in another. This aligns with the broader concept of intermodal thinking, where individuals seamlessly integrate and transfer knowledge across different modes of thought or disciplines.

 In summary, intermodal transfer is an important aspect of the learning process, and the \textit{EIS} principle provides a framework that showcases how learners progress through different modalities, facilitating the transfer of knowledge and skills between them.

\subsection{Computational Thinking}
Computational thinking is an approach that includes problem-solving and problem formulation aspects based on concepts and methods from computer science \cite{kalelioglu2016framework}. It involves breaking down complex problems into smaller parts, and then using algorithms and logical thinking to devise solutions. Computational thinking is not exclusive to computer scientists; rather, it's a fundamental skill that can be applied across various disciplines and everyday life \cite{wing2006computational}.

The principle of \textit{UMC}, which stands for \textit{Use-Modify-Create}, is often associated with computational thinking, especially in the context of learning to program, modeling, or working with software~\cite{Lylte2019}, making it more suitable for machine learning where models and similar components play a crucial role. In contrast, for example PRIMM \cite{Sue2017primm} places a stronger focus on programming.

The \textit{UMC} principle, shown in figure \ref{fig:umc}, aligns with the iterative nature of computational thinking. Individuals start by using existing solutions, move on to modifying them based on their needs or understanding, and eventually progress to creating entirely new solutions. This approach fosters a mindset of continuous improvement, adaptability, and creativity in the realm of computational problem-solving.

\begin{itemize}
    \item \textbf{Use:} When individuals \textit{use} existing algorithms or code, they are applying computational thinking to understand and leverage the functionalities provided. This can be using a programming library or software application to accomplish a specific task that involves understanding how to apply pre-existing solutions.
    \item \textbf{Modify:} \textit{Modifying} existing code or algorithms requires a deeper understanding of how they work. This process involves decomposition, pattern recognition, and algorithmic thinking, for example tweaking the parameters of an existing algorithm or adapting a piece of code for a different purpose.
    \item \textbf{Create:} \textit{Creating} original solutions, algorithms, or code is at the core of computational thinking. It involves decomposition to understand the problem, abstraction to focus on essential details, and algorithmic thinking to design a solution. This step is sometimes less strict with regards to the learners creativity and therefore can be more of a \textbf{Heavily Modify} task, which internally combines \textit{Create} and \textit{Modify} subtasks.
\end{itemize}

In educational contexts, integrating \textit{UMC} with computational thinking provides a structured approach for learners to engage with and master the principles of programming and problem-solving using computational concepts.

\subsection{Explanatory Thinking}
Explanatory thinking refers to the cognitive process of constructing and providing explanations for various phenomena, events, or concepts. It involves the ability to analyze information, identify patterns, and articulate coherent and meaningful explanations for why something happens or how it works. Explanatory thinking is a crucial aspect of critical thinking and problem-solving, allowing individuals to make sense of the world around them and communicate their understanding effectively.
In the context of \textit{AI/ML}, this would mean following a process of learning how to understand the task, working with the data and equip the students with the ability of being able to articulate (and thereby justify) modeling choices.
By reasoning about science, students get opportunities for knowledge building \cite{kenyon2019explanatory} and are more likely to attend to the interactive, invisible components in the model to explain such a process \cite{schwarz2009developing}.

\subsection{Related Tools \& Games}
Wangenheim et al. \cite{GressevonWangenheim2021} compiled visual tools for teaching \textit{ML} in the K-12 education sector and identified three key aspects of visual tools for teaching ML in K-12 education. First, they emphasize "learning by doing" through hands-on activities and model development. Second, they are based on constructivism and constructionism, promoting active knowledge construction and artifact creation. Third, they highlight the importance of adapting tools to local contexts to motivate students with relevant and interesting problems.

Exploring the use of concepts in education, Marques et al. \cite{Marques2020} examined 30 instructional units (courses, workshops, activities, etc.) primarily used in high schools in their work.
They emphasized that, given the complexity of ML concepts, several units cover only the most accessible processes.
For example, some units only teach data management, present model learning and testing at an abstract level, and hide some of the underlying \textit{ML} processes.

In~\cite{Alam2022}, Alam found that digital games could improve programming and computational thinking skills. They improve problem-solving through complex puzzles, introduce algorithmic thinking with rule-based gameplay, and require logical and strategic thinking. Additionally, these games foster creativity by allowing players to design elements. A critical aspect is the immediate feedback provided by the games, which supports iterative problem-solving, a fundamental component of computational thinking.

\section{Methodology \& Results}
In this section, we present and examine existing tools, platforms and digital games according to evaluation criteria based on the aforementioned didactic facets. We furthermore propose a process model for promoting explanatory thinking based on an industry-standard process model for \textit{Data Mining} tasks. Finally, we present our concept, \textit{ICE-T}, as a combination of three principles.

\subsection{RQ1: \textit{To what extent do existing games, digital tools and platforms for teaching machine learning implement the facets of intermodal transfer, as well as computational and explanatory thinking?}}

In order to address the didactic principles used in current games, digital tools, and learning platforms for teaching machine learning, we conducted an analysis of prominent tools, platforms, and games cited in the research of Wangenheim et al., Ashraf, and Zhou et al., as presented in table \ref{tbl:method_cmp}.

Our investigation focused on the extent and manner in which these resources implement the use-modify-create principle as a component of computational thinking.
Additionally, we examined the presence and transfer between enactive, iconic, and symbolic forms of representation.
For the explanatory facet, we examined whether machine learning algorithms, as well as data handling and the applied processes are taught. 

\begin{savenotes}
\begin{table*}[!hbt]
\begin{tabular}{|l|c|c|c|c|c|c|c|c|c|c|}
\hline
& \multicolumn{3}{c|}{Computational}&
      \multicolumn{3}{c|}{Intermodal}&
      \multicolumn{3}{c|}{Explanatory}\\
\hline
Tool/Platform/Game & use & mod. & create & en. & ico. & sym. & ML alg. & process & data \\
\hline
code.org  (AI for Oceans)\footnote{https://code.org/oceans}           
& \checkmark & & & \checkmark &   &     &  & \checkmark & \checkmark  \\

Google Teachable Machine\footnote{https://teachablemachine.withgoogle.com}\cite{Carney2020} 
& \checkmark & \checkmark  & \checkmark & \checkmark & \checkmark & &  & \checkmark & \checkmark \\

Machine Learning For Kids\footnote{https://machinelearningforkids.co.uk}             
& \checkmark & \checkmark & \checkmark & \checkmark & & \checkmark* & & \checkmark & \checkmark \\

mblock (Face recognition system)\footnote{https://planet.mblock.cc/project/234478}    
& \checkmark & \checkmark & \checkmark & & & \checkmark* & & \checkmark & \checkmark \\

Minecraft Education: Unit 9 (AI)\footnote{https://education.minecraft.net/de-de/lessons/hour-of-code-generation-ai}    
& & &\checkmark & \checkmark & &\checkmark* & & \checkmark & \checkmark \\

orange3\footnote{https://orangedatamining.com}  
& \checkmark    & \checkmark  &  \checkmark & & \checkmark & & & \checkmark & \checkmark\\

SnAIp (block-based platform)\footnote{https://snap.berkeley.edu/search?query=snaip}         
& \checkmark & \checkmark & \checkmark & & & \checkmark & \checkmark & \checkmark &   \\

Tensorflow Playground \footnote{https://playground.tensorflow.org}
& \checkmark & \checkmark  &  & \checkmark  & \checkmark &  &  & \checkmark & \checkmark \\

MIT APP Inventor\footnote{https://appinventor.mit.edu}                     
& \checkmark & \checkmark  & \checkmark &  &  & \checkmark* & \checkmark & \checkmark & \checkmark \\

While True: Learn()\footnote{https://luden.io/wtl/}                 
& & & \checkmark & \checkmark & \checkmark & & & \checkmark & \checkmark \\

\hline
\end{tabular}
\caption{Evaluation and comparison of digital tools, platforms, and games that emphasize computational thinking through use, modify, and create capabilities, intermodal transfer across enactive, iconic, and symbolic representations, and explanatory thinking using ML algorithms, processes, and data. Checkmarks signify that the respective concept plays a major role. \textit{Remark: $^*$(Model deployment)}}
\label{tbl:method_cmp}
\end{table*}
\end{savenotes}
After the analysis from Zhou et al. they recommend the use of the \textit{use-modify-create} method as a suitable approach \cite{Zhou2020}.
Our in-depth analysis shows that apart from \textit{code.org (AI for Oceans)}, \textit{Tensorflow Playgrounds} and the two games \textit{Minecraft Education (Unit 9)} and \textit{While True: Learn()}, all tools and platforms have integrated the whole concept. 

The advantage of using multiple forms of representation for explanation has not yet been widely adopted.
While \textit{Google Teachable Machine}, \textit{Tensorflow Playgrounds} and \textit{While True: Learn()} cover enactive aspects mentioned by Wangenheim \cite{GressevonWangenheim2021}, they also exploit the potential of iconic explanation.
At the symbolic representation level, many tools and platforms use the model deployment approach, which focuses on how to use a model and lets the ML-algorithm remains a black-box. 

As Marques et al. found in their analysis of workshops, courses, and activities, we also found in all the tools, games, and platforms we studied, that machine learning is always taught by focusing on the processes \cite{Marques2020}. 
Some tools focus on more accessible processes; for example, \textit{code.org} uses a training phase in which users distinguish fish from garbage, followed by a testing phase in which users watch the computer apply what it has learned.
This is followed by the decision-making process, where the user decides whether model should be trained further.
\textit{Orange 3} has a strong emphasis on the process, with a focus on workflow and the iconic representation of results, e.g. through the visualization of trees, box plots, and scatter plots.

Teaching machine learning can benefit from a combination of didactic principles.
The MIT suggests using \textit{Google´s Teachable Machine} to develop a model before using it with their \textit{APP Inventor} platform.
This combination could provide multi-faceted support for teaching ML concepts.

\subsection{RQ2: \textit{How can a process model for teaching Machine Learning be derived from a Data Mining process model?}}
To answer this research question, we introduce \textit{CRISP-DM}, before proposing a scheme based on the process model for the context of designing iterative learning units for \textit{Machine Learning} students in education. Other applications have been highlighted by Schroer et al. in~\cite{schroer2021systematic}. 
\ \\
Due to its better flexibility and universality \cite{azevedo2008kdd} compared to other process models (e.g. \textit{SEMMA}, \textit{KDD}), the \textit{CRoss Industry Standard Process for Data Mining} \cite{wirth2000crisp} appears to be most suited for adaptation into an educational context. 
This process model for \textit{Data Mining} or \textit{Data Science} tasks is widely used in industry.

\begin{figure}[!h]
\centering
\begin{minipage}{.5\textwidth}
  \centering
  \includegraphics[width=0.85\columnwidth]{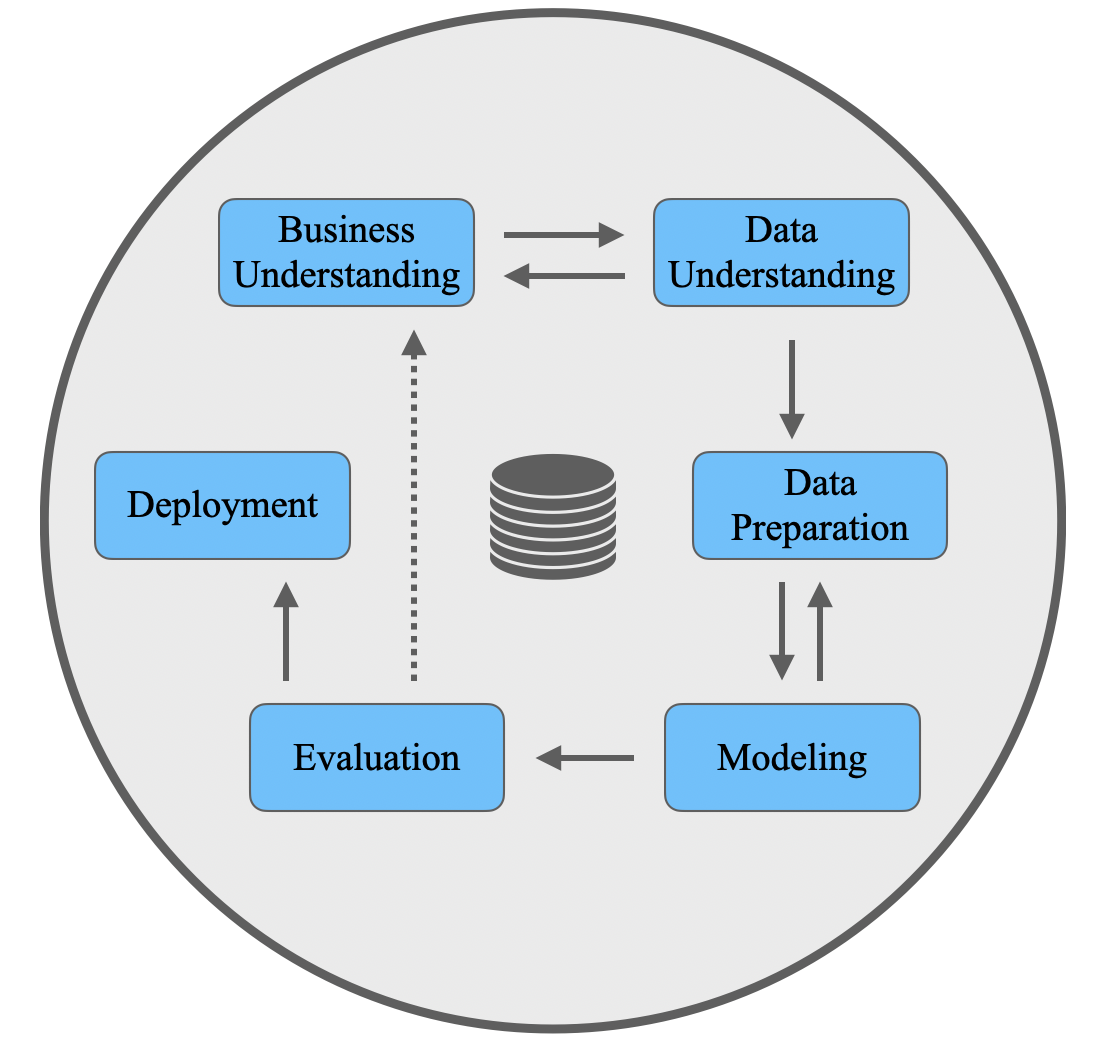}
  \caption{CRoss Industry Standard \\Process for Data Mining.}
  \label{fig:crisp-dm}
\end{minipage}%
\begin{minipage}{.5\textwidth}
  \centering
  \includegraphics[width=0.85\columnwidth]{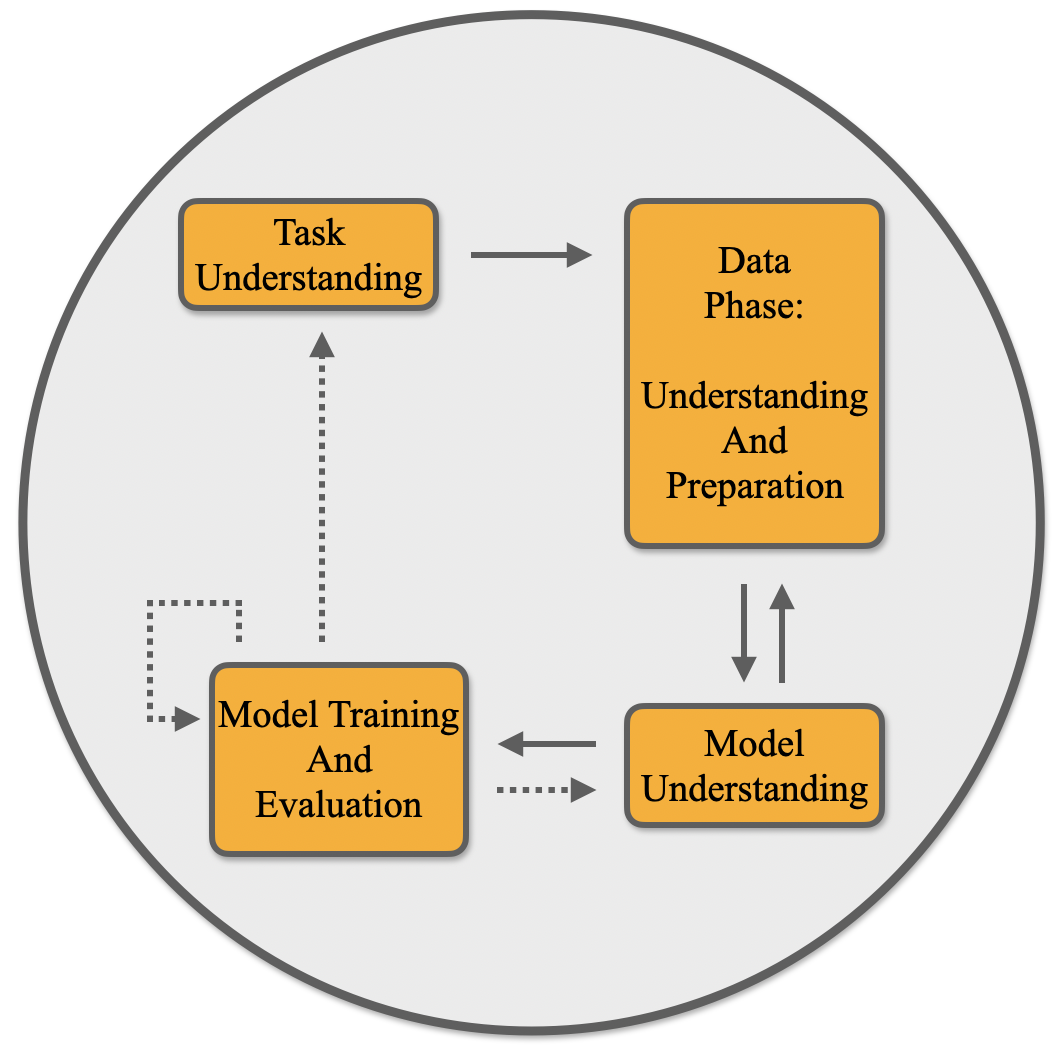}
  \caption{Promotion of Explanatory Thinking Standard Process for Machine Learning.}
  \label{fig:petsp-ai}
\end{minipage}
\end{figure}

Figure \ref{fig:crisp-dm} shows the six phases of the process: 
\begin{enumerate}
    \item \textit{Business Understanding} focuses on the requirements and objectives of a use case, which are then converted into a Data Mining (or Data Science) problem. 
    \item \textit{Data Understanding} describes a phase that consists of acquiring initial data and understanding its characteristics. From this, the goal is to identify potential issues and challenges.
    \item \textit{Data Preparation} covers the construction of the data set that will be used in the modeling phase. In this phase it can be necessary to remove, clean, engineer or transform the raw data into a format suitable for analysis.
    \item \textit{Modeling} means selecting modeling techniques and algorithms and then building predictive or descriptive models based on the prepared data. 
    \item \textit{Evaluation} is usually done during, and after the training process. Here, the models' quality gets measured and compared to the criteria set during the business understanding phase.
    \item \textit{Deployment} is the final step to integrate the \textit{Machine Learning} model into an operational environment, as an application.
\end{enumerate}
\textit{CRISP-DM} is an iterative process. Even after the deployment phase, if there are additional insights to be gained or if business objectives evolve, the process may cycle back to earlier phases, such as refining the business understanding or adjusting the data preparation and modeling steps.

Since our goal is to incorporate a principle that promotes explanatory thinking, we adapt the \textit{CRISP-DM} process model to an educational setting.
In contrast to CRISP-DM, the Business Understanding phase is intended to provide a given problem/task description. Here, concise formulation is a requirement. The next phase, \textit{Data Understanding}, can be part of the descriptive text, but can also be integrated into tasks such as exploratory analysis (e.g. read/plot data). 

Another option is to combine \textit{Data Understanding} and \textit{Data Preparation} and directly design tasks for modifying data after a given description of the data, and most likely constraints of the method that is to be applied. We denote the set of phases 2 and 3 as the \textit{Data phase}. We postulate that designing tasks for this phase is a mandatory step for learners to be able to get a deeper understanding of the data and in turn of the actual method.
The Model Understanding phase follows. To promote an understanding of \textit{AI}, we believe it is important to not treat the method/algorithm as a black-box, but instead design tasks in a way that aim to teach how these methods make choices.

Only now should the \textit{Model Training} phase start, and with it tasks that concern the actual method.
As a last phase, \textit{Evaluation} metrics are applied and the model is evaluated. Depending on the method, and due to the iterative design of the process, it is possible to revisit any of the previous phases.

The \textit{Deployment} phase plays a bigger role in practical applications (CRISP-DM), but can likely be omitted in K-12 educational cases. We note that we still believe that it has its merits if the specific goal is to support experimental learning opportunities, e.g. through group projects.

In consequence, we define the phases of the Promotion of Explanatory Thinking Standard Process for Machine Learning (\textit{short: PETSP-ML}), depicted in figure \ref{fig:petsp-ai} as such:

\begin{enumerate}
    \item \textbf{Task Understanding} can be defined as the phase where students gain a comprehensive understanding of a given academic task or assignment. This phase involves students actively engaging with the requirements, objectives, and expectations of the task to ensure clarity and alignment with their learning goals. In this context, it is especially important to convey why \textit{ML} helps with a given problem. 
    \item \textbf{Data Phase} consists of Data Understanding \& Data Preparation. Learners should have the task of dealing with the data in a practical way by visualizing the data and being able to transform it in a guided environment.
    \item \textbf{Model Understanding} refers to the phase where individuals seek to comprehend the inner workings of a machine learning model. This phase is crucial for gaining insights into how the model makes decisions, understanding its limitations, and ensuring that the model aligns with the intended goals.
    \item The \textbf{Model Training and Evaluation} phase is a crucial step where students make a model and learn from the provided data to make predictions or perform a specific task. They should gain insight into its behavior and predictions through evaluating the training. This step can also be iterated multiple times on its own if necessary.
\end{enumerate}

 \subsection{RQ3: \textit{How can the presented didactic principles be combined and applied?}}

 \begin{figure}[!hbt]
    \centering
    \includegraphics[width=0.7\columnwidth]{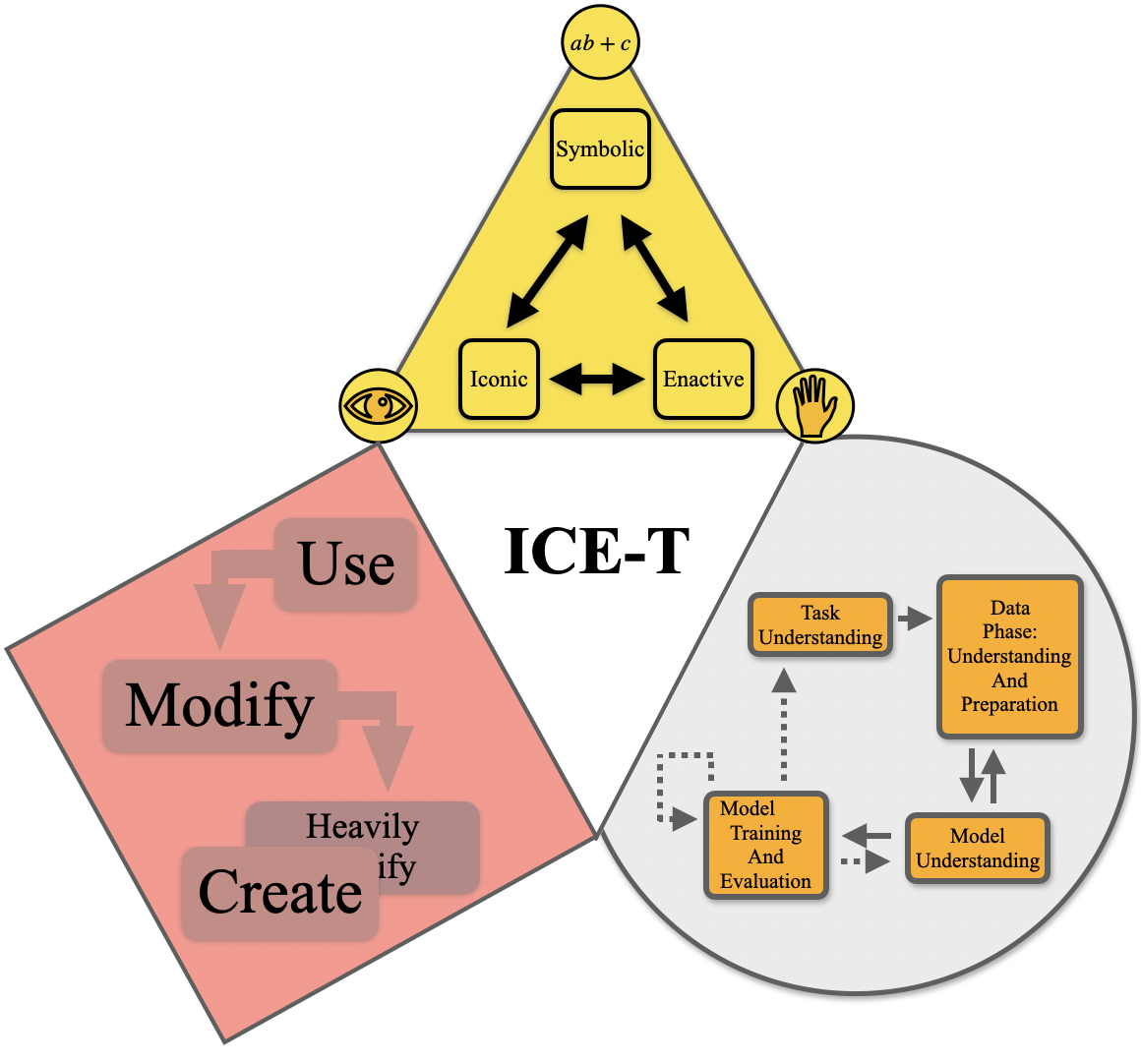}
    \caption{The three facets of the ICE-T principle: Intermodal Transfer, Computational Thinking, and Explanatory Thinking.}
    \label{fig:ice-t}
\end{figure}

\begin{figure}[!h]
    \centering
    \includegraphics[width=0.8\columnwidth]{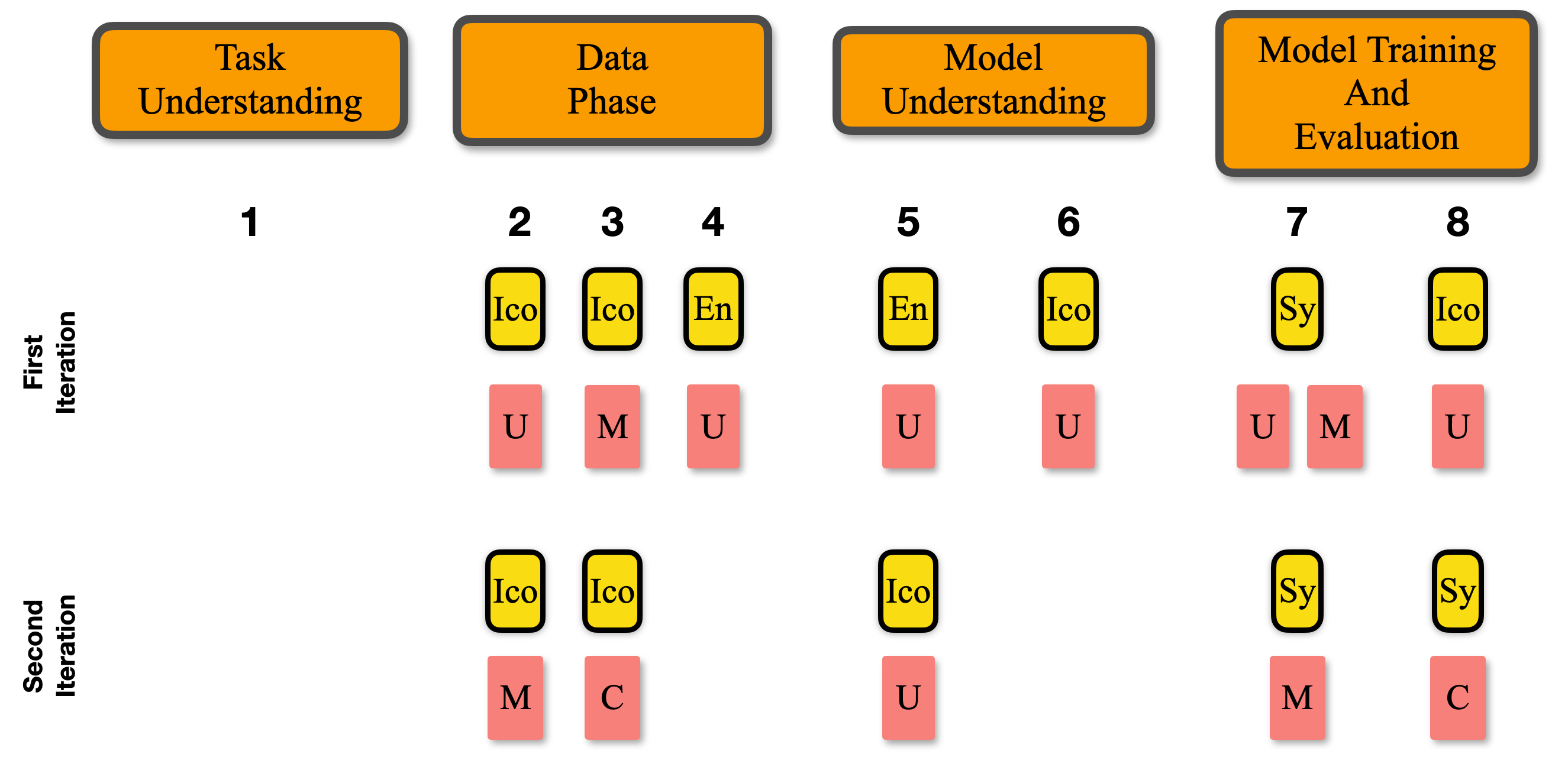}
    \caption{The iterative process of teaching the use of decision trees for animal classification is illustrated using all elements of ICE-T. The diagram shows four phases of PETSP-ML (Task Understanding, Data Phase, Model Understanding, and Model Training and Evaluation) over two iterations. Each phase indicates different activities and elements based on the EIS and UMC principles at each step.}
    \label{fig:fallbsp}
\end{figure}

We finally present our own learning concept, called \textit{ICE-T}, which combines the principles of \textit{Intermodal Transfer}, \textit{Computational Thinking} and \textit{Explanatory Thinking}, and which is illustrated in figure \ref{fig:ice-t}. We argue that in order to fully understand and use \textit{ML} methods, it is necessary to cover all of these areas when designing learning units for school curricula, learning platforms, etc.
To support our claim, we will refer to figure \ref{fig:fallbsp} and provide an example: 

We consider the situation in which an educator plans to teach how to use \textit{Decision Trees} \cite{von1980structuring} as a classifier for animals (\textbf{1:}). In this example, the data consists of a set of the animals \textit{{Lion, Shark, Eagle, Penguin}}, each with information about their attributes (\textit{Has feathers?, Can fly?, Eats meat?, Has fins?}).
The task (\textit{"Classify different animal species by asking as few questions about their characteristics as possible"}) is presented in a problem-centered manner, where students should discuss ideas to come up with solutions - independent of the algorithm that will be taught.
To understand the data, learners first examine tables of animals and their characteristics (\textbf{2:} \textit{Iconic + Use}).
Students can also do some of the preparation by expanding to the table and adjusting the data (\textbf{3:} \textit{Iconic + Modify}).
To promote a deeper understanding, they can visualize the data in different ways, e.g., by plotting (\textbf{4:} \textit{Enactive, Iconic + Use}).

To understand the model, it can be helpful to first play a "question-answer" game, in which players can decide the order in which the questions will be answered, with the goal of asking as few questions as possible in total (\textbf{5:} \textit{Enactive + Use}). Additionally, after each question, a tree-like structure could show how the question splits the data depending on the chosen question. Playing multiple rounds could also allow for comparison of resulting trees (\textbf{6:} \textit{Iconic + Use}).

For training, students could change model parameters as a programming task (\textbf{7:} \textit{Symbolic + Use and Modify}) and evaluate the immediate result according to certain rules (\textbf{8:} \textit{Iconic + Use}). Here, it might also be helpful to show the effect of trying to classify data that was not part of training the classifier.

In further iterations of the process (spiral approach), complexity should increase. 
As an example, the data could now also be generated by the students, and they should experience the algorithmic construction of the tree and model selection through the effect of splitting the data before training different models, or experiment with node splitting criteria (\textbf{second iteration:} \textit{Iconic + Symbolic + Use + Modify + Heavily Modify/Create}).

\section{Conclusion}
We examined didactic principles that get employed for teaching computer science and maths, defined didactic criteria and, based on those, evaluated a selection of existing platforms, tools and games. 
Additionally, we highlighted the issue of portraying \textit{ML} mostly as a black-box and hence the lack of focus on creating an understanding of data, algorithm and model that comes with it. To address this, we presented our multi-faceted concept, \textit{ICE-T}.

Our concept includes different representation facets, the utilization of the use-modify-create framework, and our \textit{PETSP-ML} model for educational needs based on the \textit{CRISP-DM} model.
The concept aims to provide teachers, planners of learning units, and learning platform developers with guiding didactic principles to enhance the understanding of \textit{ML} and facilitate teaching in this important area. 
Given the growing importance of \textit{ML} in today's world, it is crucial to introduce students to the fundamentals of this complex technology.

We hope that this paper will contribute to the dialogue about how to teach \textit{ML} and that planners of learning units, creators of learning platforms and educators can benefit from being guided by our concept.
\begin{credits}
\subsubsection{\ackname} \textit{This work was supported by the Federal Ministry of Education and Research of Germany and the state of North-Rhine Westphalia as part of the Lamarr-Institute for Machine Learning and Artificial Intelligence.}
\end{credits}
%
%
%
%
\bibliographystyle{splncs04}
\bibliography{literature}
\end{document}